\documentclass[journal=jacsat,manuscript=article]{achemso}

\usepackage[version=3]{mhchem} 

\usepackage{graphicx}
\usepackage{dcolumn}
\usepackage{bm}
\usepackage{mathrsfs}
\usepackage[]{xcolor}
\usepackage{siunitx}
\usepackage{float}
\usepackage{amsmath,amssymb}

\usepackage[normalem]{ulem}

\newcommand{\overlap}{\mathscr{O}}
\DeclareSIUnit\angstrom{\text {Å}}

\newcommand{\ket}[1]{\vert #1 \rangle}

\newcommand{\bra}[1]{\langle #1 \vert}

\newcommand{\Dbraket}[2]{\langle #1 \hspace{.10em} \vert \hspace{.10em}  #2 \rangle}
\newcommand{\Tbraket}[3]{\langle #1 \hspace{.10em} \vert \hspace{.10em} #2 \hspace{.10em} \vert \hspace{.10em} #3 \rangle}
\newcommand{\TBraket}[3]{\Big\langle #1 \hspace{.10em} \Big\vert \hspace{.10em} #2 \hspace{.10em} \Big\vert \hspace{.10em} #3 \Big\rangle}
\newcommand{\DBraket}[2]{\Big\langle #1 \hspace{.10em} \Big\vert \hspace{.10em}  #2 \Big\rangle}

\newcommand{\hf}{\mathrm{HF}}

\renewcommand{\phi}{\varphi}

\newcommand{\transp}{^\mathrm{T}}

\title{Coupled cluster theory for nonadiabatic dynamics: nuclear gradients and nonadiabatic couplings in similarity constrained coupled cluster theory}

\author{Eirik F.~Kjønstad}
\email{eirik.kjonstad@ntnu.no}
\affiliation{%
 Department of Chemistry, Norwegian University of Science and Technology, 7491 Trondheim,
Norway
}%
 \alsoaffiliation{Department of Chemistry, Stanford University, Stanford, CA, USA}
 \alsoaffiliation{Stanford PULSE Institute, SLAC National Accelerator Laboratory,
Menlo Park, CA, USA}
\author{Sara Angelico}
\affiliation{%
 Department of Chemistry, Norwegian University of Science and Technology, 7491 Trondheim,
Norway
}
\author{Henrik Koch}
\email{henrik.koch@ntnu.no}
\affiliation{%
 Department of Chemistry, Norwegian University of Science and Technology, 7491 Trondheim,
Norway
}

\date{\today}

\begin{document}

\begin{abstract}
Coupled cluster theory is one of the most accurate electronic structure methods for predicting ground and excited state chemistry. However, the presence of numerical artifacts at electronic degeneracies, such as complex energies,  has made it difficult to apply it in nonadiabatic dynamics simulations.
While it has already been shown that such numerical artifacts can  be fully removed by using  similarity constrained coupled cluster (SCC) theory [\emph{J.~Phys.~Chem.~Lett.}~2017,~\textbf{8},~19,~4801–4807], simulating dynamics requires efficient implementations of gradients and nonadiabatic couplings. Here, we present an  implementation of nuclear gradients and nonadiabatic derivative couplings at the similarity constrained coupled cluster singles and doubles (SCCSD) level of theory, thereby making possible nonadiabatic dynamics simulations using a coupled cluster theory that provides a correct description of conical intersections between excited states. We present a few numerical examples that show good agreement with literature values and discuss some limitations of the method. 
\end{abstract}

\maketitle

\section{Introduction}
Upon photoexcitation to an excited electronic state, molecular systems normally undergo relaxation through one or more conical intersections. When a system approaches such electronic degeneracies, the dynamics of the nuclei changes: the nuclear motion must then be treated in terms of a nuclear wavepacket and the coupling between the electrons and the nuclei (nonadiabatic coupling) becomes large, leading to a breakdown of the Born-Oppenheimer approximation. As a result, non-radiative transfer of nuclear population between electronic states becomes possible, something that typically occurs within tens or hundreds of femtoseconds after excitation by light.\citep{domcke2004conical}

It is challenging to simulate such a process, however, as it requires an accurate treatment not only of the nuclear wavepacket but also of the electronic states involved. The choice of the electronic structure method often qualitatively alters the predicted dynamics, emphasizing the importance of a balanced and accurate treatment of the electronic states.\citep{mai2020molecular} Depending on the system of interest, it can be important to effectively capture static correlation (for example, by using complete active space methods\citep{roos1980complete}) or dynamical correlation (for example, by using perturbation theory\citep{Andersson1990} or density functional theory\citep{huix2020time, Li2014, Hennefarth2023}). In the latter category, coupled cluster theory\citep{koch1990coupled,stanton1993equation} is often particularly accurate,\citep{loos2020} making it a promising candidate for simulating photochemical processes.

A major theoretical issue has hindered this, however.
Coupled cluster methods are known to produce numerical artifacts at same-symmetry 
intersections, where the potential energy surfaces become distorted and the energies can become complex-valued. This fact led some authors\citep{hattig2005structure,kohn2007can} to question whether the method could  be used to simulate excited state dynamics. As was recently shown by the authors and collaborators, however, the numerical artifacts of the method can be removed by constraining the electronic states to be orthogonal, which furthermore implies a correct intersection dimensionality (two directions lift the degeneracy) and a proper first-order behavior as the degeneracy is lifted (the intersections are conical).\citep{kjonstad2017crossing, kjonstad2017resolving, kjonstad2019orbital} These developments indicated that the modified coupled cluster method, known as similarity constrained coupled cluster (SCC)\citep{kjonstad2017resolving, kjonstad2019orbital} theory, would be applicable in excited state nonadiabatic dynamics simulations.

Large-scale simulations require efficient implementations of the nuclear energy gradients and the nonadiabatic derivative coupling elements. In the case of coupled cluster energy gradients, both derivation techniques and efficient implementations are well-established and widespread.\citep{handy1984evaluation, helgaker1988analytical, koch1990derivatives, stanton1994} Less attention has been given to the nonadiabatic couplings. There have been two recent implementations at the coupled cluster singles and doubles level of theory (CCSD), although these arrive at the coupling elements through two different routes: one evaluates the couplings in terms of summed-state gradients\citep{Tajti2009,faraji2018calculations} while the other (implemented by the authors) directly evaluates them by using standard Lagrangian/Z-vector techniques.\citep{christiansen1999first, kjonstad2023coupling} There is some uncertainty about whether these are equivalent.\citep{kjonstad2023coupling}

In this paper, we derive and implement analytical derivative couplings and energy gradients for the similarity constrained coupled cluster singles and doubles (SCCSD)\citep{kjonstad2019orbital} method, building on our recent CCSD implementations of analytical derivative coupling elements\citep{kjonstad2023coupling} and nuclear energy gradients.\citep{schnack2022} Here we will focus on 
implementation aspects. 
In a recent study on thymine, we applied the implementation to perform the first nonadiabatic dynamics simulations with CCSD and SCCSD.\citep{kjonstad2024thymine}

\section{Theory}

\subsection{Coupled cluster theory}
In coupled cluster theory, 
the ground state wave function is written as\cite{coester1958bound,bartlett2007coupled}
\begin{align}
    \vert \psi_0 \rangle = \exp(\mathscr{T}) \vert \mathrm{HF} \rangle, \label{eq:cc_ansatz}
\end{align}
where $\mathscr{T}$ is called the cluster operator and the reference wave function $\vert \mathrm{HF} \rangle$ is usually taken to be the Hartree-Fock state. By applying the exponential operator, the mean-field wave function $\vert \mathrm{HF} \rangle$ is transformed into a correlated wave function $\vert \psi_0 \rangle$ that more accurately represents the electronic ground state.

In practice, the cluster operator  
incorporates excitations only up to a given rank. For instance, it may include single and double excitations with respect to $\vert \mathrm{HF} \rangle$, in which case we obtain the coupled cluster with singles and doubles (CCSD)\citep{purvis1982full} method. More generally, the cluster operator is written
\begin{align}
    \mathscr{T} = \sum_{\mu \in \mathscr{E}_e} t_\mu \tau_\mu = T_1 + T_2 + \ldots + T_n,
\end{align}
where $\mathscr{E}_e$ denotes the included excitations, the subscript $k$ in $T_k$ denotes the excitation rank, and the sum truncates at a given excitation order $n$. For example, $\mathscr{T} = T_1 + T_2$ in CCSD. Each excitation operator $\tau_\mu$ is weighted by an amplitude $t_\mu$, and this amplitude indicates the weight in $\vert \psi_0 \rangle$ of the associated configuration $\vert \mu \rangle = \tau_\mu \vert \mathrm{HF} \rangle$.

These cluster amplitudes are determined by projecting the Schrödinger equation onto the excitation subspace $\mathscr{E}_e$. We start from the Schrödinger equation for $\vert \psi_0 \rangle$,
\begin{align}
    H \vert \psi_0 \rangle = E_0 \vert \psi_0 \rangle, \label{eq:SE}
\end{align}
where $H$ denotes the electronic Hamiltonian, $\vert \psi_0 \rangle$ is given in Eq.~\eqref{eq:cc_ansatz}, and $E_0$ is the ground state energy. Next, we project the Schrödinger equation onto  
\begin{align}
\begin{split}
    \{ \langle \mathrm{HF} \vert \exp(-\mathscr{T}), \langle \mu \vert \exp(-\mathscr{T}) \}, 
\end{split} \label{eq:proj_basis}
\end{align}
where $\mu \in \mathscr{E}_e$.  This projection procedure gives a set of equations that determine $E_0$ and $t_\mu$:
\begin{align}
    E_0 &= \langle \mathrm{HF} \vert \bar{H} \vert \mathrm{HF} \rangle \\
    0 &= \langle \mu \vert \bar{H} \vert \mathrm{HF} \rangle = \Omega_\mu,
\end{align}
where 
\begin{align}
        \bar{H} = \exp(-\mathscr{T}) H \exp(\mathscr{T}). \label{eq:H-bar-without-kappa}
\end{align}
If we want to make the orbital dependence explicit, we can also write\citep{hald2003lagrangian}
\begin{align}
    \bar{H} = \exp(-\mathscr{T}) \exp(\kappa) H \exp(-\kappa) \exp(\mathscr{T}), \label{eq:H-bar-with-kappa}
\end{align}
where the orbital rotation operator is given by
\begin{align}
    \kappa = \sum_{ai} \kappa_{ai} (E_{ai} - E_{ia}) = \sum_{ai} \kappa_{ai} E_{ai}^-.
\end{align}
Here, $E_{ai}$ is the singlet excitation operator from the occupied orbital $i$ to the virtual orbital $a$.
In calculations at a specific geometry, $\kappa = 0$ is normally understood to correspond to the Hartree-Fock orbitals, and we write $\bar{H}$ as in Eq.~\eqref{eq:H-bar-without-kappa}. When we evaluate nuclear gradients, we need the latter form, Eq.~\eqref{eq:H-bar-with-kappa}, as $\kappa$ will be different from zero when we make displacements away from the geometry where we wish to evaluate the nuclear gradient.

A peculiar feature of coupled cluster theory, which stems from the non-Hermiticity of $\bar{H}$, is the introduction of left (or bra) states that are different from the right (or ket) states. In particular, $\vert \psi_0 \rangle^\dagger \neq \langle \psi_0 \vert$. The left ground state does not have an exponential form, like in Eq.~\eqref{eq:cc_ansatz}, but is instead expressed as
\begin{align}
    \bra{\psi_0} = \Bigl( \bra{\mathrm{HF}} + \sum_{\mu \in \mathscr{E}_e} \bar{t}_\mu \bra{\mu} \Bigr) \exp(-\mathscr{T}). \label{eq:psi0_left}
\end{align}
Nonetheless, like $\ket{\psi_0}$, it is also determined by projecting the Schrödinger equation. Starting from 
\begin{align}
    \bra{\psi_0} H = \bra{\psi_0} E_0,
\end{align}
we project onto 
\begin{align}
    \{ \exp(\mathscr{T}) \ket{\mathrm{HF}}, \exp(\mathscr{T}) \ket{\mu} \},
\end{align}
where $\mu \in \mathscr{E}_e$.
This results in an equation for $\bar{t}_\mu$:\citep{helgaker2013molecular}
\begin{align}
    0 = \Tbraket{\hf}{[\bar{H},\tau_\mu]}{\hf} + \sum_{\nu \in \mathscr{E}_e} \bar{t}_\nu \Tbraket{\nu}{[\bar{H},\tau_\mu]}{\hf}.
\end{align}

The excited states can be described with two alternative but closely related approaches, the linear response\citep{koch1990coupled} and equation of motion theories.\citep{stanton1993equation} In the equation of motion approach, which we adopt here, the excited states are expressed as
\begin{align}
    \ket{\psi_n} &= \exp(\mathscr{T}) \Bigl( R_0^n \ket{\hf} + \sum_{\mu \in \mathscr{E}_e} R_\mu^n \ket{\mu} \Bigr) \label{eq:eom_R} \\
    \bra{\psi_n} &= \Bigl( L_0^n \bra{\hf} + \sum_{\mu \in \mathscr{E}_e} L_\mu^n \bra{\mu} \Bigr) \exp(-\mathscr{T}),\label{eq:eom_L}
\end{align}
and these states are determined by the same projection procedure used
for the ground state. It will be convenient to define $\ket{0} = \ket{\hf}$ so that $\mu = 0$ corresponds to the reference state and $\tau_0 = 1$. Then, if we let $\mathscr{E}$ denote the set containing both $\mu = 0$ and $\mu \in \mathscr{E}_e$,  we can write
\begin{align}
    \ket{\psi_n} &= \exp(\mathscr{T}) \Bigl( \sum_{\mu \in \mathscr{E}} R_\mu^n \ket{\mu} \Bigr)   \\
    \bra{\psi_n} &= \Bigl( \sum_{\mu \in \mathscr{E}} L_\mu^n \bra{\mu} \Bigr) \exp(-\mathscr{T}).
\end{align}

When we repeat the ground state projection procedure for the excited states, we find that the excited state amplitudes satisfy the eigenvalue equations
\begin{align}
    \bar{\boldsymbol{\mathcal{H}}} \boldsymbol{\mathcal{R}}_n &= E_n \boldsymbol{\mathcal{R}}_n \label{eq:eom-eigenvalue-R} \\
    \boldsymbol{\mathcal{L}}_n\transp \bar{\boldsymbol{\mathcal{H}}}  &= E_n \boldsymbol{\mathcal{L}}_n\transp,
\end{align}
where
\begin{align}
   \bar{\mathcal{H}}_{\mu\nu} = \Tbraket{\mu}{\bar{H}}{\nu}, \quad \mu,\nu \in \mathscr{E},
\end{align}
i.e.,
\begin{align}
    \bar{\boldsymbol{\mathcal{H}}} = \begin{pmatrix}
        E_0 & \boldsymbol{\eta}^T \\
        \boldsymbol{0}   & \bar{\boldsymbol{H}}
    \end{pmatrix}, \quad \eta_\nu = \Tbraket{\mathrm{HF}}{\bar{H}}{\nu}, \quad \bar{H}_{\mu\nu} = \Tbraket{\mu}{\bar{H}}{\nu}, \quad \mu,\nu \in \mathscr{E}_e,
\end{align}
and
\begin{align}
    \boldsymbol{\mathcal{R}}_n = 
    \begin{pmatrix}
        R_n^0 \\ \boldsymbol{R}_n
    \end{pmatrix}, \quad
    \boldsymbol{\mathcal{L}}_n\transp = 
    \begin{pmatrix}
        L_n^0 \; \boldsymbol{L}_n\transp
    \end{pmatrix}.
\end{align}
Here, $E_n$, with $n = 1,2,\ldots$, denotes the excited state energy. Also the ground state wave functions satisfy the same eigenvalue equations, with
\begin{align}
    \boldsymbol{\mathcal{R}}_0 = 
    \begin{pmatrix}
        1 \\ \boldsymbol{0}
    \end{pmatrix}, \quad \boldsymbol{\mathcal{L}}_0\transp = 
    \begin{pmatrix}
        1 \; \bar{\boldsymbol{t}}\transp
    \end{pmatrix}.
\end{align}

Once $\bra{\psi_n}$ and $\ket{\psi_n}$ have been determined, 
we can evaluate the energy gradient
as
\begin{align}
    \boldsymbol{g}_n = \nabla \Tbraket{\psi_n}{H}{\psi_n}
\end{align}
and the nonadiabatic derivative coupling vector as 
\begin{align}
    \boldsymbol{d}_{mn} = \Dbraket{\psi_m}{\nabla \psi_n}.
\end{align}
These two quantities, $\boldsymbol{g}_n$ and $\boldsymbol{d}_{mn}$, are of special interest in molecular dynamics. The gradient $\boldsymbol{g}_n$ provides the force acting on the nuclei, while the derivative coupling $\boldsymbol{d}_{mn}$ is responsible for nonadiabatic transitions between electronic states. 
 
The derivative coupling vector diverges at conical intersections, where the electronic states are degenerate. This is easily seen in the exact limit, where we have 
\begin{align}
    \boldsymbol{d}_{mn} = \Dbraket{\psi_m}{\nabla \psi_n} = \frac{\Dbraket{\psi_m}{(\nabla H) \psi_n}}{E_m - E_n}.
\end{align}
A proper description of such intersections is therefore essential if a method is to be applied in photochemical applications. Somewhat surprisingly, it turns out\citep{kjonstad2017crossing,kjonstad2017resolving,kjonstad2019orbital} that these intersections can only be described correctly if the method guarantees some sort of orthogonality between the excited states, something which is \emph{not} true in standard coupled cluster theory.

\subsection{Similarity constrained coupled cluster theory}

In similarity constrained coupled cluster theory, which was designed for applications to nonadiabatic dynamics,\citep{kjonstad2017crossing,kjonstad2017resolving,kjonstad2019orbital} a set of electronic states are constrained to be orthogonal. Here, a subset  of states $\mathscr{K}$ are required to satisfy
\begin{align}
    \mathscr{O}_{kl} = \Tbraket{\psi_k}{\mathscr{P}}{\psi_l} = 0, \quad k \neq l \in \mathscr{K}, \label{eq:orthogonality-general}
\end{align}
where $\mathscr{P}$ is a projection operator and so $\mathscr{O}_{kl}$ is an approximation of $\Dbraket{\psi_k}{\psi_l}$.
To enforce the orthogonality of the states, the cluster operator is expressed as
\begin{align}
    \mathscr{T} = \sum_{\mu \in \mathscr{E}} t_\mu \tau_\mu + \sum_{k\neq l \in \mathscr{K}} \zeta_{kl} X_{kl},
\end{align}
where each $X_{kl}$ is an excitation operator whose rank is higher than the $\mu$ included in $\mathscr{E}$, and  $\zeta_{kl}$ enforces $\mathscr{O}_{kl} = 0$. 
Both $\mathscr{P}$ and $X_{kl}$ can be chosen in several different ways, leading to different variants of the theory.\citep{kjonstad2017resolving,kjonstad2019orbital} Here we will assume that they only depend on the right excited states, so that
\begin{align}
    \mathscr{P} = \mathscr{P}(\{\mathcal{R}_n\}), \quad X_{kl} = X_{kl}(\mathcal{R}_k, \mathcal{R}_l),
\end{align}
as this is required for the method to scale correctly with the size of the system (that is, to be size-extensive/size-intensive).\citep{kjonstad2019orbital} The similarity constrained coupled cluster method can be considered an extension of the standard coupled cluster method, since all the standard equations are kept unchanged. The only changes to the method are the additional
orthogonality conditions
$\mathscr{O}_{kl} = 0$ and the additional excitation operators in $\mathscr{T}$. 
Note, however, that the orthogonality conditions couple the ground state to the excited states in $\mathscr{K}$, implying that the ground and excited states must be determined simultaneously.

\subsection{Nuclear gradients and derivative coupling elements}

\subsubsection{Lagrangians}
As noted already, the nuclear gradients and derivative couplings are the two main components needed to simulate nonadiabatic excited state   dynamics. We therefore need to know how to evaluate these elements efficiently in our approximate electronic structure methods. The generally adopted approach is to apply the Langragian method, also called the Z-vector method, which allows us to effectively evaluate gradients (for example, of the energy) by imposing a set of constraints on the parameters of the electronic structure method.\citep{handy1984evaluation,helgaker1988analytical}

Its main advantage here is that it removes the need to explicitly evaluate the derivatives of these parameters, for example, $\mathrm{d}t_\mu/\mathrm{d}x$, where $x$ is a nuclear coordinate. These derivatives are instead accounted for through the constraints and the associated Lagrangian multipliers. 
If we assume that we want to evaluate the gradient of $E$ with respect to $\boldsymbol{x}$, where $E$ depends on a set of parameters $\boldsymbol{\lambda}$ that can be determined by solving a set of equations $\boldsymbol{e} = 0$, then the Lagrangian takes the form
\begin{align}
    \mathscr{L}(\boldsymbol{\lambda}, \bar{\boldsymbol{\lambda}}, \boldsymbol{x}) = E(\boldsymbol{\lambda}, \boldsymbol{x}) + \sum_{p} \bar{\lambda}_{p} e_p(\boldsymbol{\lambda}, \boldsymbol{x}),
\end{align}
where $\bar{\boldsymbol{\lambda}}$ contains the so-called Lagrangian multipliers.
The constraints ($\boldsymbol{e} = 0$) are imposed by making the Lagrangian stationary with respect to the multipliers $\bar{\boldsymbol{\lambda}}$, while the multipliers are determined by making the Lagrangian stationary with respect to the parameters $\boldsymbol{\lambda}$. In other words, we have
\begin{align}
    \frac{\partial \mathscr{L}}{\partial \lambda_p} = 0, \quad \frac{\partial \mathscr{L}}{\partial \bar{\lambda}_p} = e_p = 0,
\end{align}
which implies that
\begin{align}
    \frac{\mathrm{d}E}{\mathrm{d}x} = \frac{\mathrm{d}\mathscr{L}}{\mathrm{d}x} = \frac{\partial\mathscr{L}}{\partial x} + \sum_p \frac{\partial\mathscr{L}}{\partial \lambda_p} \frac{\partial\lambda_p}{\partial x} + \sum_p \frac{\partial\mathscr{L}}{\partial \bar{\lambda}_p} \frac{\partial\bar{\lambda}_p}{\partial x} = \frac{\partial\mathscr{L}}{\partial x}.
\end{align}
The \emph{total} derivative of $E$ is replaced by the \emph{partial} derivative of $\mathscr{L}$, and the latter contains no derivative of the parameters with respect to $\boldsymbol{x}$.

Our constraints are simply the equations that determine the parameters; for example, one set of constraints are the ground state equations $\Omega_\mu = 0$, since these equations determine the amplitudes $t_\mu$. Each set of constraints is given a set of Lagrangian multipliers; for example, the ground state equations $\Omega_\mu = 0$ are associated with a set of multipliers $\bar{\zeta}_\mu$. Written out explicitly,
the nuclear gradient $\boldsymbol{g}_n$, where $n \in \mathscr{K}$, can be evaluated from the Lagrangian
\begin{align}
\begin{split}
    \mathscr{L}_n &= E_n   
    + \sum_{k \in \mathscr{K}} \bar{\boldsymbol{\mathcal{L}}}_k\transp (\bar{\boldsymbol{\mathcal{H}}}-E_k)\boldsymbol{\mathcal{R}}_k
    \\
    &+ \sum_{k \in \mathscr{K}} \bar{E}_k (1 - \Dbraket{\Lambda_k}{R_k}) + \sum_{k \neq l \in \mathscr{K}} \bar{\gamma}_{kl} \mathscr{O}_{kl} + \sum_{\mu} \bar{\zeta}_\mu \Omega_\mu + \sum_{ai} \bar{\kappa}_{ai} F_{ai},
\end{split} \label{eq:Lagrangian_gradient}
\end{align}
where $F$ denotes the Fock matrix, $a$ and $i$ denote a virtual and an occupied orbital, respectively, and the energy of the $k$th state is expressed as
\begin{align}
E_k = \Tbraket{\Lambda_k}{\bar{H}}{R_k}. \label{eq:energy-lambda}
\end{align}
Here, $\bra{\Lambda_k}$ is some (for now unspecified) vector that is normalized with respect to $\ket{R_k}$, that is, $\Dbraket{\Lambda_k}{R_k} = 1$. 
Note that Eq.~\eqref{eq:energy-lambda} will then be equal to the energy as a result of the eigenvalue equation in Eq.~\eqref{eq:eom-eigenvalue-R}.
The first term in $\mathscr{L}_n$ is the energy, and the rest are the Lagrangian constraints. The first and second constraints ensure that the states in $\mathscr{K}$ are  
eigenstates and that $E_k$ is the energy of the $k$th state, while the third ensures that the states in $\mathscr{K}$ are orthogonal (that is, $\mathscr{O}_{kl} = 0$ for $k \neq l \in \mathscr{K}$). The final two constraints are the ground state coupled cluster equations and the Hartree-Fock equations, which respectively determine the cluster amplitudes $t_\mu$ and the molecular orbitals.

The Lagrangian for the derivative couplings is identical to that for the nuclear gradient (because the constraints $\boldsymbol{e}=0$ are the same), except for the first term. The coupling $\boldsymbol{d}_{mn}$ between $m, n\in \mathscr{K}$ can be evaluated from 
\begin{align}
\begin{split}
    \mathscr{L}_{mn} &= O_{mn} 
    + \sum_{k \in \mathscr{K}} \bar{\boldsymbol{\mathcal{L}}}_k\transp (\bar{\boldsymbol{\mathcal{H}}}-E_k)\boldsymbol{\mathcal{R}}_k
    \\
    &+ \sum_{k \in \mathscr{K}} \bar{E}_k (1 - \Dbraket{\Lambda_k}{R_k}) + \sum_{k \neq l \in \mathscr{K}} \bar{\gamma}_{kl} \mathscr{O}_{kl} + \sum_{\mu} \bar{\zeta}_\mu \Omega_\mu + \sum_{ai} \bar{\kappa}_{ai} F_{ai},
\end{split} \label{eq:Lagrangian_coupling}
\end{align}
where, as proposed by Hohenstein,\citep{hohenstein2016analytic} we 
define the bra-frozen overlap
\begin{align}
    O_{mn} = \Dbraket{\psi_m(\boldsymbol{x}_0)}{\psi_n(\boldsymbol{x})}.
\end{align}
Here, $O_{mn}$ introduces a parametric dependence on some reference geometry $\boldsymbol{x}_0$. This means that this Lagrangian can be used to evaluate the coupling exactly at $\boldsymbol{x}_0$ and not at any other value of $\boldsymbol{x}$. However, since $\boldsymbol{x}_0$ can be chosen freely, this implies no loss of generality.

\subsubsection{Expressions for the gradient and couplings}
Once the multipliers are known, we can evaluate the gradients and couplings as
\begin{align}
    \boldsymbol{g}_n &= \nabla E_n =  \frac{\mathrm{d}E_n}{\mathrm{d}\boldsymbol{x}} = \frac{\partial \mathscr{L}_n}{\partial \boldsymbol{x}}  
\end{align}
and
\begin{align}
    \boldsymbol{d}_{mn} &= \Dbraket{\psi_m}{\nabla \psi_n} = \DBraket{\psi_m}{\frac{\mathrm{d}\psi_n}{\mathrm{d}\boldsymbol{x}}} = \frac{\partial \mathscr{L}_{mn}}{\partial \boldsymbol{x}}.
\end{align}
Let us first write out the gradient expression in detail. Note that when we take the partial derivative of the Lagrangian, we can treat the parameters as constants as their dependence on the nuclear coordinates is included in the multiplier terms. In addition, the dependence of the creation and annihilation operators on $\boldsymbol{x}$ can be ignored, see Ref.~\citenum{helgaker1988analytical}. Thus, the only explicit nuclear dependence resides in the Hamiltonian integrals.
We will denote the nuclear derivatives $\{ \partial H / \partial x_q \vert_0 \}$ as $H^{(1)}$ and let
\begin{align}
        \bar{H}^{(1)} = \exp(-\mathscr{T}) \exp(\kappa) H^{(1)} \exp(-\kappa) \exp(\mathscr{T}),
\end{align}
where $\kappa = 0$ corresponds to the Hartree-Fock orbitals at $\boldsymbol{x}_0$. 
Here, the Hamiltonian $H$ is expressed in terms of a set of orthonormal molecular orbitals (OMOs) that are related to the so-called unmodified MOs or UMOs (defined as the MOs obtained from the MO coefficients at $\boldsymbol{x}_0$ and the AOs at $\boldsymbol{x}$) through an orbital connection matrix. In the case of the symmetric connection, which we adopt here, we have
\begin{align}
    H^{(1)} = H^{[1]} - \frac{1}{2} \{ S^{[1]}, H \}, \label{eq:reortho}
\end{align}
where $H^{[1]}$ and $S^{[1]}$ refer to the derivative of the UMO Hamiltonian and  overlap, and $\{ A, B \}$ denotes the operator obtained by one-index transformations of $A$ and $B$. 
 The latter term in Eq.~\eqref{eq:reortho} leads to the  ``reorthonormalization terms'' in the gradient. We refer to the literature for more details on orbital connections.\citep{olsen1995orbital}

Since the derivative only acts on the Hamiltonian, it is convenient to introduce some further notation. In particular, if we let
\begin{align}
\begin{split}
  \label{eq:K-gradients}
        \bra{K_k} &= (\delta_{kn} - \Dbraket{\bar{\mathcal{L}}_k}{\mathcal{R}_k}) \bra{\Lambda_k} + \bra{\bar{L}_k},
\end{split} 
\end{align}
the Lagrangian becomes
\begin{align}
\begin{split}
        \mathscr{L}_n &= \sum_{k \in \mathscr{K}} (\Tbraket{K_k}{\bar{H}}{R_k} + \bar{L}_0^k \Tbraket{\mathrm{HF}}{\bar{H}}{\mathcal{R}_k}) \\
        &\quad+ \Tbraket{\bar{\zeta}}{\bar{H}}{\hf} + \sum_{ai} \bar{\kappa}_{ai} F_{ai} \\
        &\quad+ \sum_{k \in \mathscr{K}} \bar{E}_k (1 - \Dbraket{\Lambda_k}{R_k}) + \sum_{k \neq l \in \mathscr{K}} \bar{\gamma}_{kl} \overlap_{kl}
\end{split}
\end{align}
and the nuclear gradient becomes
\begin{align}
\begin{split}
        \boldsymbol{g}_n &= \mathscr{L}_n^{(1)} = \sum_{k \in \mathscr{K}} ( \Tbraket{K_k}{\bar{H}^{(1)}}{R_k} + \bar{L}_0^k \Tbraket{\mathrm{HF}}{\bar{H}^{(1)}}{\mathcal{R}_k} ) \\
        &\quad\quad\quad\quad\quad\quad+ \Tbraket{\bar{\zeta}}{\bar{H}^{(1)}}{\hf} + \sum_{ai} \bar{\kappa}_{ai} F_{ai}^{(1)}. \label{eq:gradient_expression}
\end{split}
\end{align}
Note that, when we apply the derivative, the last two terms in $\mathscr{L}_n$ vanish because there is no $H$-dependence in these terms and so the partial derivatives are zero. The expression for $\boldsymbol{g}_n$ found here is similar to the standard coupled cluster case,\citep{schnack2022} except that there are additional excited state terms which appear due to the coupling of states in $\mathscr{K}$ that arises due to the orthogonality conditions (in this case, the $\bra{K_k}$ and $\bar{L}^k_0$ terms with $k \neq n$).

The expression for the derivative coupling is similar. In particular, we find that
\begin{align}
\begin{split}
    \boldsymbol{d}_{mn} = \mathscr{L}_{mn}^{(1)} &= O_{mn}^{(1)} + \sum_{k \in \mathscr{K}} ( \Tbraket{K_k}{\bar{H}^{(1)}}{R_k} + \bar{L}_0^k \Tbraket{\mathrm{HF}}{\bar{H}^{(1)}}{\mathcal{R}_k} ) \\
        &\quad\quad+ \Tbraket{\bar{\zeta}}{\bar{H}^{(1)}}{\hf} + \sum_{ai} \bar{\kappa}_{ai} F_{ai}^{(1)},
\end{split}    
    \label{eq:coupling_expression}
\end{align}
where we instead have 
\begin{align}
    \label{eq:K-couplings}
    \bra{K_k} = - \Dbraket{\bar{\mathcal{L}}_k}{\mathcal{R}_k} \bra{\Lambda_k} + \bra{\bar{L}_k}.
\end{align}

Both $\boldsymbol{g}_n$ and $\boldsymbol{d}_{mn}$ consist of several terms that are transition elements of the gradient Hamiltonian $\bar{H}^{(1)}$. These are conveniently calculated in terms of density matrices; for general $\bra{\mathcal{L}}$ and $\ket{\mathcal{R}}$, we have
\begin{align}
    \Tbraket{\mathcal{L}}{\bar{H}^{(1)}}{\mathcal{R}} = \sum_{pq} h_{pq}^{(1)} d_{pq}(\boldsymbol{\mathcal{L}}, \boldsymbol{\mathcal{R}}) + \frac{1}{2}\sum_{pqrs} g_{pqrs}^{(1)} d_{pqrs}(\boldsymbol{\mathcal{L}}, \boldsymbol{\mathcal{R}}) + \Dbraket{\mathcal{L}}{\mathcal{R}} h_\mathrm{nuc}^{(1)},
\end{align}
where
\begin{align}
    d_{pq}(\boldsymbol{\mathcal{L}}, \boldsymbol{\mathcal{R}}) &= \Tbraket{\mathcal{L}}{\exp(-\mathscr{T})E_{pq}\exp(\mathscr{T})}{\mathcal{R}} \\
    d_{pqrs}(\boldsymbol{\mathcal{L}}, \boldsymbol{\mathcal{R}}) &= \Tbraket{\mathcal{L}}{\exp(-\mathscr{T})e_{pqrs}\exp(\mathscr{T})}{\mathcal{R}}, \quad e_{pqrs} = E_{pq}E_{rs} - \delta_{qr} E_{ps},
\end{align}
and where we have used that the Hamiltonian can be expressed as
\begin{align}
    H = \sum_{pq} h_{pq} E_{pq} + \frac{1}{2} \sum_{pqrs} g_{pqrs} e_{pqrs} + h_\mathrm{nuc}.
\end{align}
In the above equations, $h_{pq}$ and $g_{pqrs}$ denote the one- and two-electron Hamiltonian integrals, respectively, and $h_{\mathrm{nuc}}$ denotes the nuclear repulsion energy. The indices $p,q,r,$ and $s$ denote general molecular orbitals.

For both the gradient and the coupling elements, we evaluate the reorthonormalization terms by substituting $H^{(1)}$ in Eq.~\eqref{eq:reortho} into $\boldsymbol{g}_n$ and $\boldsymbol{d}_{mn}$ and collecting the contributions that stem from $\{S^{[1]}, H\}$. The end-result can be written as a contraction of $S^{[1]}$ with a generalized Fock matrix. These terms are identical in CCSD and SCCSD (albeit with modified density matrices) and we refer to the literature for the detailed expressions.\citep{hald2003lagrangian,schnackpetersen2022}

As in other electronic structure methods, $\boldsymbol{d}_{mn}$ can be subdivided into two terms, one that can induce transitions by translational motion (by violating the sum rule\citep{Subotnik2011}) and one that does not induce such transitions. The term in $\boldsymbol{d}_{mn}$ that violates the sum rule is $O_{mn}^{(1)}$, which can be expressed as
\begin{align}
    O_{mn}^{(1)} = \frac{1}{2} \sum_{pq} d_{pq}(\mathcal{L}_m, \mathcal{R}_n) (W^{(1)}_{pq}-W^{(1)}_{qp}),
\end{align}
where
\begin{align}
    W^{(1)}_{pq} = \sum_{\alpha\beta} C_{\alpha p} C_{\beta q} \Dbraket{\chi_\alpha}{\chi_\beta^{(1)}}.
\end{align}
Here, $C_{\alpha p}, C_{\beta q}$ denote the molecular orbital coefficients and $\chi_{\alpha}, \chi_\beta$ denote atomic orbitals.
For a derivation of this expression for $O_{mn}^{(1)}$, see, e.g., Ref.~\citenum{kjonstad2023coupling}.

\subsubsection{Stationarity equations: nuclear gradient}
Evaluating the gradient and couplings is only possible after we have determined the Lagrangian multipliers. Let us begin with the gradient Lagrangian $\mathscr{L}_n$. The multipliers are determined from the stationarity equations, which involve derivatives with respect to the orbital rotation parameters, $\kappa_{ai}$, the cluster amplitudes, $t_\mu$, and the excited state amplitudes, $R_\mu^k$ and $\zeta_{kl}$.

It turns out that the stationarity equations have a simpler form if the lambda states, $\bra{\Lambda_k}$, at $\boldsymbol{x}_0$, are chosen to be equal to the left excited state vectors, that is, if
\begin{align}
    \bra{\Lambda_k} = \bra{L_k},
\end{align}
and we will assume this in the remainder of the text.
Let us start by considering stationarity with respect to $\bra{\Lambda_m}$, which will allow us to determine the multipliers $\bar{E}_m$. In fact, since
\begin{align}
    \frac{\partial \mathscr{L}_n}{\partial \Lambda_\mu^m}\Big\vert_0 = 0 = R_\mu^m \Big( E_m(\delta_{mn} - \Dbraket{\mathcal{\bar{L}}_m}{\mathcal{R}_m}) - \bar{E}_m \Big),
\end{align}
we can ensure stationarity by setting
\begin{align}
    \bar{E}_m = E_m (\delta_{mn} - \Dbraket{\mathcal{\bar{L}}_m}{\mathcal{R}_m}).
\end{align}
In addition, it turns out that the multipliers relating to the Hartree-Fock contribution to
the excited states, that is, $\bar{L}^k_0$, can be expressed in terms of the orthogonality multipliers $\bar{\gamma}_{kl}$. In particular, we have
\begin{align}
    \frac{\partial \mathscr{L}_n}{\partial R_0^m}\Big\vert_0 = 0 = - \omega_m \bar{L}_0^m + \sum_{k \neq l \in \mathscr{K}} \bar{\gamma}_{kl} \frac{\partial \mathscr{O}_{kl}}{\partial R_0^m}, 
\end{align}
where $\omega_m = E_m - E_0$, and, therefore,
\begin{align}
    \bar{L}_0^m = \sum_{k \neq l \in \mathscr{K}} \bar{\gamma}_{kl} \Bigl( \frac{1}{\omega_m} \frac{\partial \mathscr{O}_{kl}}{\partial R_0^m} \Bigr).
\end{align}

The remaining multipliers ($\boldsymbol{\bar{\zeta}}, \boldsymbol{\bar{L}}_k,\bar{\gamma}_{kl}, \boldsymbol{\bar{\kappa}}$) are non-redundant and are determined by
a linear system of equations that can be solved with standard techniques, for example, the Davidson algorithm.\citep{DAVIDSON197587} But we can make one more simplification to this set of equations. The orbital multipliers ($\boldsymbol{\bar{\kappa}}$) depend on the ground and excited state multipliers ($\boldsymbol{\bar{\zeta}}, \boldsymbol{\bar{L}}_k, \bar{\gamma}_{kl}$), but not the other way around. Hence, we can determine the ground and excited state multipliers without first knowing the orbital multipliers. 

Let us therefore first consider the response equations for the ground and excited state multipliers ($\boldsymbol{\bar{\zeta}}, \boldsymbol{\bar{L}}_k, \bar{\gamma}_{kl}$). For the cluster amplitude stationarity, we find that 
\begin{align}
\begin{split}
    \frac{\partial \mathscr{L}_n}{\partial t_\mu}\Big\vert_0 &= 0 = \sum_\nu L^n_\nu \bar{F}(\boldsymbol{R}_n)_{\nu\mu} + \sum_\nu \bar{\zeta}_\nu A_{\nu\mu} + \sum_{k \in \mathscr{K}} \sum_{\nu\tau} \bar{L}^k_\nu P^k_{\nu\tau}  \bar{F}(\boldsymbol{R}_k)_{\tau\mu} \\
    &+ \sum_{l \neq j \in \mathscr{K}} \bar{\gamma}_{lj} \Big\{ \frac{\partial \mathscr{O}_{lj}}{\partial t_\mu} + \sum_{k \in \mathscr{K}}\frac{1}{\omega_k} \frac{\partial \mathscr{O}_{lj}}{\partial R^k_0} \Big( R^k_0 \big( \eta_\mu - \sum_\nu L^k_\nu \bar{F}(\boldsymbol{R}_k)_{\nu\mu} \big) + \sum_\nu R^k_\nu F^0_{\nu\mu} \Big) \Big\},
\label{eq:t-response-gradient}
\end{split}
\end{align}
where we have defined the various $\boldsymbol{t}$-derivative terms
\begin{align}
    &A_{\nu\mu} = \Tbraket{\nu}{[\bar{H},\tau_\mu]}{\mathrm{HF}} = \Tbraket{\nu}{\bar{H}}{\mu} - E_0 \\ 
    &\bar{F}(\boldsymbol{R}_k)_{\nu\mu} = \Tbraket{\nu}{[\bar{H},\tau_\mu]}{R_k} \\ 
    &\eta_\mu = \Tbraket{\mathrm{HF}}{[\bar{H}, \tau_\mu]}{\mathrm{HF}} \\
    &F^0_{\nu\mu} = \Tbraket{\mathrm{HF}}{[[\bar{H},\tau_\nu],\tau_\mu]}{\mathrm{HF}},
\end{align}
as well as a projection matrix that removes components along the $k$th excited state:
\begin{align}
    P^k_{\nu\tau} = \delta_{\nu\tau} - R^k_\nu L^k_\tau.
\end{align}
In matrix notation, the stationarity equation can be written
\begin{align}
\begin{split}
    \boldsymbol{0} &= \boldsymbol{L}_n\transp \bar{\boldsymbol{F}}(\boldsymbol{R}_n) + \bar{\boldsymbol{\zeta}}\transp\boldsymbol{A} + \sum_{k \in \mathscr{K}} \bar{\boldsymbol{L}}_k\transp \boldsymbol{P}_k \bar{\boldsymbol{F}}(\boldsymbol{R}_k) \\
    &+ \sum_{l\neq j \in \mathscr{K}} \bar{\gamma}_{lj} \Big\{ \nabla_t \mathscr{O}_{lj} + \sum_{k \in \mathscr{K}} \frac{1}{\omega_k} \frac{\partial \mathscr{O}_{lj}}{\partial R^k_0} \Big( R^k_0 ( \boldsymbol{\eta}\transp - \boldsymbol{L}_k\transp \bar{\boldsymbol{F}}(\boldsymbol{R}_k) )  + \boldsymbol{R}_k\transp \boldsymbol{F}_0  \Big) \Big\}.
\end{split} \label{eq:t-response-gradient-vectornot}
\end{align}
Note that Eq.~\eqref{eq:t-response-gradient-vectornot} consists of four terms, the first being a constant term, since it does not depend on any of the multipliers, while the remaining three terms are linear in $\boldsymbol{\bar{\zeta}}, \boldsymbol{\bar{L}}_k$, and $\bar{\gamma}_{kl}$, respectively. This will also be the case for the two next stationarity conditions.

For the excited state amplitude stationarity, we find that
\begin{align}
\begin{split}
    &\frac{\partial \mathscr{L}_n}{\partial R^m_\mu}\Big\vert_0 = 0 = \sum_\nu L^n_\nu Y^m(\boldsymbol{R}_n)_{\nu\mu} + \sum_\nu \bar{\zeta}_\nu Y^m_{\nu\mu} \\ 
    &+ \sum_{k \in \mathscr{K}} \sum_\nu \bar{L}^k_\nu \Big\{ \delta_{km}  (A_{\nu\mu} - \omega_m \delta_{\nu\mu})  
    + \sum_\tau P^k_{\nu\tau} Y^m(\boldsymbol{R}_k)_{\tau\mu} \Big\} \\
    &+ \sum_{l \neq j \in \mathscr{K}} \bar{\gamma}_{lj} \Big\{ \frac{\partial \mathscr{O}_{lj}}{\partial R^m_\mu} + \sum_{k \in \mathscr{K}} \frac{1}{\omega_k} \frac{\partial \mathscr{O}_{lj}}{\partial R^k_0} \Big(\delta_{km} \eta_\mu +  R^k_0 \big( Y^{m,0}_\mu - \sum_\nu L^k_\nu Y^m(\boldsymbol{R}_k)_{\nu\mu}\big) + Y^{m,0}(\boldsymbol{R}_k)_{\mu} \Big) \Big\},
\end{split}
\label{eq:R-response-gradient}
\end{align}
where we have defined the various $\boldsymbol{R}_m$-derivative terms
\begin{align}
    &Y^m(\boldsymbol{R}_k)_{\nu\mu} = \Tbraket{\nu}{Y^m_\mu}{R_k}, \quad Y^m_\mu = \frac{\partial \bar{H}}{\partial R^m_\mu}, \label{eq:Y-first} \\
    &Y^m_{\nu\mu} = \Tbraket{\nu}{Y^m_\mu}{\mathrm{HF}} \\
    &Y^{m,0}_\mu = \Tbraket{\mathrm{HF}}{Y_\mu^m}{\mathrm{HF}} \\
    &Y^{m,0}(\boldsymbol{R}_k)_\mu = \Tbraket{\mathrm{HF}}{Y_\mu^m}{R_k}.
\end{align}
In matrix notation, this stationarity equation can be written
\begin{align}
\begin{split}
    \boldsymbol{0} &= \boldsymbol{L}_n\transp \boldsymbol{Y}_m(\boldsymbol{R}_n) + \bar{\boldsymbol{\zeta}}\transp \boldsymbol{Y}_m + \sum_{k \in \mathscr{K}} \bar{\boldsymbol{L}}_k\transp\bigl\{ \delta_{km} ( \boldsymbol{A} - \omega_m ) 
    + \boldsymbol{P}_k \boldsymbol{Y}_m(\boldsymbol{R}_k) \bigr\} \\
    &+ \sum_{l \neq j \in \mathscr{K}} \bar{\gamma}_{lj} \Big\{ \nabla_{R_m} \mathscr{O}_{lj} + \sum_{k \in \mathscr{K}} \frac{1}{\omega_k} \frac{\partial \mathscr{O}_{lj}}{\partial R^k_0} \Big( \delta_{km}\boldsymbol{\eta}\transp +  R^k_0 \big( \boldsymbol{Y}_{m,0}\transp - \boldsymbol{L}_k\transp \boldsymbol{Y}_m(\boldsymbol{R}_k) \big) + \boldsymbol{Y}_{m,0}(\boldsymbol{R}_k)\transp \Big) \Bigr\}.
\end{split} \label{eq:R-response-gradient-vectornot}
\end{align}
Finally, for $\zeta_{kl}$, we find 
\begin{align}
\begin{split}
    \frac{\partial \mathscr{L}_n}{\partial \zeta_{kl}}\Big\vert_0 &= 0 = \sum_{\mu\nu} L^n_\mu Z^{kl}_{\mu\nu} R^n_\nu + \sum_\mu \bar{\zeta}_\mu z^{kl}_{\mu} + \sum_{j \in \mathscr{K}} \sum_{\mu\nu\tau} \bar{L}^j_{\mu} 
    P^j_{\mu\tau} Z^{kl}_{\tau\nu} R^j_\nu \\ 
    &+ \sum_{i \neq p \in \mathscr{K}} \bar{\gamma}_{ip} \Big\{ \sum_{j \in \mathscr{K}}  \frac{1}{\omega_j} \frac{\partial \mathscr{O}_{ip}}{\partial R^j_0} \Big( R_0^j \big( z_{kl} - \sum_{\mu\nu} L^j_{\mu} Z^{kl}_{\mu\nu} R^j_\nu \big) + \tilde{z}^{kl}_{\mu} R_\mu^j  \Big) \Big\},
\end{split}
\label{eq:zeta-response-gradient}
\end{align}
where
\begin{align}
    Z^{kl}_{\mu\nu} = \TBraket{\mu}{\frac{\partial \bar{H}}{\partial \zeta_{kl}}}{\nu}, \quad 
    z^{kl}_{\mu} = \TBraket{\mu}{\frac{\partial \bar{H}}{\partial \zeta_{kl}}}{\mathrm{HF}}, \quad 
    \tilde{z}^{kl}_{\nu} = \TBraket{\mathrm{HF}}{\frac{\partial \bar{H}}{\partial \zeta_{kl}}}{\nu}, \quad z_{kl} = \TBraket{\mathrm{HF}}{\frac{\partial \bar{H}}{\partial \zeta_{kl}}}{\mathrm{HF}}
\end{align}
In matrix notation, this stationarity equation can be written
\begin{align}
\begin{split}
    \boldsymbol{0} &= \boldsymbol{L}_n\transp \boldsymbol{Z}_{kl} \boldsymbol{R}_n + \bar{\boldsymbol{\zeta}}\transp \boldsymbol{z}_{kl} + \sum_{j \in \mathscr{K}} \bar{\boldsymbol{L}}_j\transp \boldsymbol{P}_j 
    \boldsymbol{Z}_{kl} \boldsymbol{R}_j \\
    &\quad\quad\quad+ \sum_{i \neq p \in \mathscr{K}} \bar{\gamma}_{ip} \Big\{ \sum_{j \in \mathscr{K}} \frac{1}{\omega_j} \frac{\partial \mathscr{O}_{ip}}{\partial R^j_0} \Big( R^j_0 ( z_{kl} - \boldsymbol{L}_j\transp \boldsymbol{Z}_{kl} \boldsymbol{R}_j ) + \tilde{\boldsymbol{z}}_{kl}\transp\boldsymbol{R}_j  \Big) \Bigr\}.
\end{split} \label{eq:zeta-response-gradient-vectornot}
\end{align}

This concludes our presentation of the ground and excited state response equations. To summarize, Eqs.~\eqref{eq:t-response-gradient-vectornot}, \eqref{eq:R-response-gradient-vectornot}, and \eqref{eq:zeta-response-gradient-vectornot} form a coupled set of linear equations that can be solved using standard techniques. The next step is to determine the orbital multipliers.

In the case of orbital stationarity, we find that
\begin{align}
    \frac{\partial \mathscr{L}_n}{\partial \kappa_{ai}}\Big\vert_0 = 0 = \eta^\kappa_{ai} + \sum_{bj} \bar{\kappa}_{bj} A_{bjai}^\mathrm{HF},
\end{align}
where
\begin{align}
    A_{bjai}^\mathrm{HF} = \frac{\partial F_{bj}}{\partial \kappa_{ai}}\Big\vert_0,
\end{align}
and where
\begin{align}
    \eta^\kappa_{ai} &=  \sum_{k \in \mathscr{K}} \Bigl(  \Tbraket{K_k}{\bar{H}^\kappa_{ai}}{R_k} + \bar{L}^k_0 \Tbraket{\mathrm{HF}}{\bar{H}^\kappa_{ai}}{\mathcal{R}_k}\Bigr) + \Tbraket{\bar{\zeta}}{\bar{H}^\kappa_{ai}}{\mathrm{HF}},
\end{align}
with
\begin{align}
    \bar{H}^\kappa_{ai} &= \frac{\partial \bar{H}}{\partial \kappa_{ai}}\Big\vert_0 = \exp(-\mathscr{T})[E_{ai}^-, H]\exp(\mathscr{T}).
\end{align}
Expressions for $\boldsymbol{A}^\mathrm{HF}$ can be found elsewhere.\citep{schnackpetersen2022} The terms in $\boldsymbol{\eta}_\kappa$ can be expressed in terms of densities. In fact, for general $\bra{\mathcal{L}}$ and $\ket{\mathcal{R}}$, we have\citep{hald2003lagrangian}
\begin{align}
\begin{split}
    &\Tbraket{\mathcal{L}}{\bar{H}_{ai}^\kappa}{\mathcal{R}} = -\mathscr{P}_{ai}^- \Bigl( \sum_t h_{at} \bigl(d_{ti}(\mathcal{L},\mathcal{R}) + d_{it}(\mathcal{L},\mathcal{R})\bigr) \\
    &\quad\quad+ \sum_{trs} g_{atrs} \bigl( d_{itrs}(\mathcal{L},\mathcal{R}) + d_{tirs}(\mathcal{L},\mathcal{R}) \bigr) \Bigr),
\end{split}
\end{align}
where
\begin{align}
    \mathscr{P}_{ai}^- X_{ai} = X_{ai} - X_{ia}.
\end{align}

\subsubsection{Stationarity equations: derivative coupling}
Looking at the gradient and coupling Lagrangians,  $\mathscr{L}_n$ and $\mathscr{L}_{mn}$, we see that they only differ in
the first term, which is equal to $E_n$ in $\mathscr{L}_n$ and $O_{mn}$ in $\mathscr{L}_{mn}$; see Eqs.~\eqref{eq:Lagrangian_gradient} and \eqref{eq:Lagrangian_coupling}. 
As a result, the differences in the stationarity equations only stem from this first term. The other terms are identical.
Let us start by writing out the bra-frozen overlap:
\begin{align}
\begin{split}
    O_{mn} &= \Dbraket{\psi_m(\boldsymbol{x}_0)}{\psi_n(\boldsymbol{x})} \\
    &=\Tbraket{\mathcal{L}_m}{\exp(-\mathscr{T})\exp(\kappa)\big\vert_0 \exp(-\kappa)\exp(\mathscr{T})}{\mathcal{R}_n}.
\end{split}
\end{align}
In the case of the ground state amplitudes, we find that
\begin{align}
    \frac{\partial O_{mn}}{\partial t_\mu}\Big\vert_0 = \Tbraket{\mathcal{L}_m}{\tau_\mu}{\mathcal{R}_n} = L^m_\mu R_0^n + \Tbraket{L_m}{\tau_\mu}{R_n}
\end{align}
Similarly, for the excited state amplitudes, we have
\begin{align}
    \frac{\partial O_{mn}}{\partial R^k_0}\Big\vert_0 = \delta_{kn} \Dbraket{\mathcal{L}_m}{\mathrm{HF}} = 0,
\end{align}
and
\begin{align}
    \frac{\partial O_{mn}}{\partial R^k_\mu}\Big\vert_0 &= \delta_{kn} L^m_\mu + \sum_{l \neq j \in \mathscr{K}} \zeta_{lj} \TBraket{\mathcal{L}_m}{\frac{\partial X_{lj}}{\partial R^k_\mu}}{\mathcal{R}_n} \\
    \frac{\partial O_{mn}}{\partial \zeta_{lj}}\Big\vert_0 &= \Tbraket{\mathcal{L}_m}{X_{lj}}{\mathcal{R}_n}.
\end{align}
Finally, for the orbital rotation parameters, we find that
\begin{align}
    \frac{\partial O_{mn}}{\partial \kappa_{ai}}\Big\vert_0 = - \Tbraket{\mathcal{L}_m}{\exp(-\mathscr{T}) E_{ai}^- \exp(\mathscr{T})}{\mathcal{R}_n} = -\mathscr{P}_{ai}^- d_{ai}(\mathcal{L}_m, \mathcal{R}_n).
\end{align}
More details about the stationarity equations for the coupling are given in Supporting Information S1.

As the stationarity equations for the coupling are identical to the ones for the gradient, except for the  constant term, the same algorithm can be used to solve for the multipliers. The only required modifications consist in
appropriately modifying the constant term, which arises from partial derivatives either of $E_n$ (for the gradient) or $O_{mn}$ (for the coupling), as well as appropriately choosing the definition of $\bra{K_k}$, see Eqs. \eqref{eq:K-gradients} and \eqref{eq:K-couplings}. Furthermore, once the multipliers have been determined, the nuclear derivatives in the coupling can also be evaluated by the same implementation as the gradient, with the exception of $O_{mn}^{(1)}$.

\subsection{Similarity constrained coupled cluster singles and doubles method}
The expressions derived so far apply to any level of theory for the similarity constrained coupled cluster method. In the remainder of the paper, we will focus on the specifics at the singles and doubles level of theory.
\subsubsection{Method}
For the similarity constrained coupled cluster singles and doubles method, we define the cluster operator as
\begin{align}
    \mathscr{T} = T_1 + T_2 + \sum_{k \neq l \in \mathscr{K}} \zeta_{kl} X_3^{kl},
\end{align}
where
\begin{align}
    X_3^{kl} = \mathscr{P}_{kl}^- R^k_1 R^l_2 = R_1^k R_2^l - R_1^l R_2^k
\end{align}
and where we have defined the one- and two-electron state excitation operators
\begin{align}
    R^k_1 = \sum_{\mu_1} R^k_{\mu_1} \tau_{\mu_1}, \quad R^k_2 = \sum_{\mu_2} R^k_{\mu_2} \tau_{\mu_2}.
\end{align}
Together with a choice of projection operator $\mathscr{P}$ in the orthogonality relations, see Eq.~\eqref{eq:orthogonality-general}, this defines the similarity constrained singles and doubles (SCCSD) method.\citep{kjonstad2019orbital} 

Several choices of $\mathscr{P}$ are possible. The most obvious choice is the one we will refer to as the natural projection, where we simply project onto the excitation subspace $\mathscr{E}$,
\begin{align}
    \mathscr{P} = \sum_{\mu \in \mathscr{E}} \ket{\mu}\bra{\mu}.
\end{align}
In the case of SCCSD, this operator will project onto the subspace defined by the Hartree-Fock reference as well as all single and double excitations out of this reference. While this choice of $\mathscr{P}$ yields the correct untruncated limit, where $\mathscr{P} = \mathbb{I}$, it also gives rise to non-zero changes in the excitation energies of non-interacting subsystems, i.e., it is not fully size-intensive.\citep{kjonstad2019orbital} We  use the shorthand SCCSD($\mathscr{E}$)  for this choice of $\mathscr{P}$. 

Another choice is what we will refer to as the state projection, where we project onto the subspace of states $\mathscr{K}$,
\begin{align}
    \mathscr{P} = \sum_{k,l \in \mathscr{K}} \ket{\mathcal{R}_k}\Dbraket{\mathcal{R}_k}{\mathcal{R}_l}^{-1}\bra{\mathcal{R}_l}.
\end{align}
In this case, we preserve size-intensivity but at the cost of projecting onto a smaller subspace. The correct limit is still formally satisfied, however, since $\mathscr{P} = \mathbb{I}$ once all states are included in $\mathscr{K}$. We use the shorthand  SCCSD($\mathscr{K}$) for this second choice of $\mathscr{P}$. 

As we will show below, the choice of projection appears to often have a small impact on the obtained gradients and derivative coupling elements. For a description of other possible projection operators, see Supporting Information S2.

\subsubsection{Implementation} 
Here we discuss the main aspects of the implementation. For a more detailed description, including programmable expressions, we refer the reader to Supporting Information S3--S5. The following discussion is meant to provide a general overview of how the gradients and coupling elements can be implemented, starting from an existing CCSD implementation. 

Our implementation is restricted to two states, and so we will assume that $\mathscr{K} = \{ a, b \}$ and suppress the subindices ``$ab$'', writing $\zeta$ instead of $\zeta_{ab}$. The cluster operator can then be expressed as 
\begin{align}
    \mathscr{T} = T_1 + T_2 + \zeta X_3, \quad X_3 = \mathscr{P}_{ab}^- R^a_1 R^b_2.
\end{align}
To relate the CCSD and SCCSD implementations, it is particularly useful to note that
\begin{align}
\begin{split}
    \bar{H} &= \exp(-\mathscr{T})H_\kappa \exp(\mathscr{T}) \\
    &= \exp(-\zeta X_3) \bar{H}_T \exp(\zeta X_3) \\
    &= \bar{H}_T + \zeta [\hat{H},  X_3] + \ldots
\end{split}
\end{align}
where
\begin{align}
    H_\kappa &= \exp(\kappa) H \exp(-\kappa) \\
    \bar{H}_T &= \exp(-T) H_\kappa \exp(T) \\
    \hat{H} &= \exp(-T_1) H_\kappa \exp(T_1).
\end{align}
Here, $\bar{H}_T$ is the similarity transformed Hamiltonian at the CCSD level of theory, and $\hat{H}$ is the so-called $T_1$-transformed Hamiltonian. The higher-order commutators do not contribute to any equations because of their high excitation rank. The SCCSD corrections arise solely from  $\zeta [\hat{H},  X_3]$.

These corrections have already been implemented for the ground and excited state equations, and we refer the reader to the original SCCSD paper, where we also provide expressions for the orthogonality conditions.\citep{kjonstad2019orbital} Here, we will only consider the changes that are required for the nuclear energy gradient and the derivative couplings, and these can be subdivided into two categories: the density matrices and the response vectors.

In the case of the density matrices, we need to evaluate the corrections
\begin{align}
    \Delta d_{pq}(\mathcal{L},\mathcal{R}) &= \zeta \Tbraket{\mathcal{L}}{[E_{pq}, X_3]}{\mathcal{R}} \\
    \Delta d_{pqrs}(\mathcal{L},\mathcal{R}) &= \zeta \Tbraket{\mathcal{L}}{[e_{pqrs}, X_3]}{\mathcal{R}},
\end{align}
where we 
can again apply rank considerations. Since $[E_{pq},X_3]$ is at least a double excitation, we 
only have one non-zero block:
\begin{align}
    d_{ia}(\mathcal{L},\mathcal{R}) = \zeta \Tbraket{L_2}{[E_{ia},X_3]}{\mathrm{HF}} R_0.
\end{align}
The same reasoning for the two-electron density leads to three distinct non-zero blocks:
\begin{align}
    \Delta d_{ijka}(\mathcal{L},\mathcal{R}) &= \zeta \Tbraket{L_2}{[e_{ijka}, X_3]}{\mathrm{HF}}R_0 \\
    \Delta d_{bcka}(\mathcal{L},\mathcal{R}) &= \zeta \Tbraket{L_2}{[e_{bcka}, X_3]}{\mathrm{HF}}R_0 \\
    \Delta d_{iajb}(\mathcal{L},\mathcal{R}) &= \zeta \big( \Tbraket{L_1}{[e_{iajb}, X_3]}{\mathrm{HF}}R_0 +  \Tbraket{L_2}{[e_{iajb}, X_3]}{R_1} \big).
\end{align}
Expressions for these density terms are given in Supporting Information S3.

The corrections to the response vectors arise because of the $R_k$ and $\zeta$-dependence of $X_3$, as well as the orthogonality conditions. For example, in the case of the $R_k$-dependence, we need to consider the $Y^k_\mu$ operator, see Eq.~\eqref{eq:Y-first}, which for SCCSD becomes
\begin{align}
    Y^k_\mu = \frac{\partial \bar{H}}{\partial R^k_\mu} = \zeta \Bigl[\hat{H}, \frac{\partial X_3}{\partial R^k_\mu} \Bigr] + \ldots,
\end{align}
where we again find that the higher-order commutators do not contribute to any of the equations. By inserting the leading term of $Y^k_\mu$, and focusing on matrix transformations (that is, the action of matrices on  vectors), we find that
\begin{align}
    \sigma^m(\boldsymbol{R}_k)_\mu = \sum_\nu L_\nu Y^m(\boldsymbol{R}_k)_{\nu\mu} &= \Tbraket{L}{Y_\mu^m}{R_k} = \Tbraket{L_2}{Y_\mu^m}{R^k_1},
\end{align}
where we have used rank-considerations to simplify: $[\hat{H}, \partial X_3/\partial R^k_\mu ]$ consists of single excitations and higher as the derivative of $X_3$ is a triple excitation operator and $\hat{H}$ is a two-electron operator. Similarly, we find that
\begin{align}
    \sigma^m_\mu = \sum_\nu L_\nu Y^m_{\nu\mu} = \Tbraket{L}{Y_\mu^m}{\mathrm{HF}} = \Tbraket{L_1}{Y_\mu^m}{\mathrm{HF}} + \Tbraket{L_2}{Y_\mu^m}{\mathrm{HF}}
\end{align}
and
\begin{align}
    Y_\mu^{m,0} &= \Tbraket{\mathrm{HF}}{Y_\mu^m}{\mathrm{HF}} = 0 \\
    Y^{m,0}(\boldsymbol{R}_k)_\mu &=  \Tbraket{\mathrm{HF}}{Y_\mu^m}{R_k} = 0.
\end{align}
Note that the $\boldsymbol{R}_k$-dependence in the the bra-frozen overlap derivative similarly vanishes due to rank considerations, in particular,
\begin{align}
    \frac{\partial O_{mn}}{\partial R^k_\mu}\Big\vert_0 &= \delta_{kn} L^m_\mu + \zeta \TBraket{\mathcal{L}_m}{\frac{\partial X_3}{\partial R^k_\mu}}{\mathcal{R}_n} = \delta_{kn} L^m_\mu. \\
    \frac{\partial O_{mn}}{\partial \zeta}\Big\vert_0 &= \Tbraket{\mathcal{L}_m}{X_3}{\mathcal{R}_n} = 0 
\end{align}
For more details on the $\boldsymbol{R}_k$-derivative terms, see Supporting Information S4.

Considering the $\zeta$-dependence of $X_3$, we find that we can reuse $Y^k_\mu$ terms:
\begin{align}
\begin{split}
   \sum_{\mu\nu} L_\mu Z_{\mu\nu} R_\nu &= \TBraket{L}{\frac{\partial \bar{H}}{\partial \zeta}}{R} \\
   &= \Tbraket{L}{[\hat{H},X_3]}{R} \\
   &= \zeta^{-1} \sum_\nu \Tbraket{L}{Y^k_\nu}{R} R^k_\nu \\
   &= \zeta^{-1} \sum_\nu \sigma^k(\boldsymbol{R})_\nu R_\nu^k,
\end{split}
\end{align}
and, similarly,
\begin{align}
\begin{split}
    \sum_\nu L_\nu z_\nu &= \TBraket{L}{\frac{\partial \bar{H}}{\partial \zeta}}{\mathrm{HF}} \\
    &= \Tbraket{L}{[\hat{H},X_3]}{\mathrm{HF}} \\
    &= \zeta^{-1} \sum_\nu \Tbraket{L}{Y^k_\nu}{\mathrm{HF}} R^k_\nu \\ 
    &= \zeta^{-1} \sum_\nu \sigma^k_\nu R_\nu^k.
\end{split}
\end{align}
Finally, the reference-to-excited components vanish due to rank-considerations:
\begin{align}
    \sum_\nu \tilde{z}_\nu R_\nu = \TBraket{\mathrm{HF}}{\frac{\partial \bar{H}}{\partial \zeta}}{R} = \Tbraket{\mathrm{HF}}{[\hat{H},X_3]}{R} = 0.
\end{align}

The orthogonalities also contribute to the response vectors, as we need to evaluate the derivatives of these conditions with respect to $T$, $R_k$, and $R^k_0$. For this purpose, it is convenient to write out the orthogonality relation, where we take SCCSD($\mathscr{E}$) as an example:
\begin{align}
    \mathscr{O}_{ab} = \Tbraket{\psi_a}{\mathscr{P}}{\psi_b} = \sum_{\mu\nu\tau \in \mathscr{E}} \mathcal{R}^a_\mu \Tbraket{\mu}{\exp(T^\dagger)}{\nu}\Tbraket{\nu}{\exp(T)}{\tau}\mathcal{R}^b_\tau = \boldsymbol{\mathcal{R}}_a\transp \boldsymbol{\mathcal{Q}}\transp \boldsymbol{\mathcal{Q}} \boldsymbol{\mathcal{R}}_b.
\end{align}
Note that we use $T$ here instead of $\mathscr{T}$; since there are no non-zero contributions to $\boldsymbol{\mathcal{Q}}$ that arise from the triple excitation in $\mathscr{T}$.
Now, 
\begin{align}
    \frac{\partial \mathscr{O}_{ab}}{\partial \mathcal{R}^b_\mu} = (\boldsymbol{\mathcal{R}}_a\transp \boldsymbol{\mathcal{Q}}\transp \boldsymbol{\mathcal{Q}})_\mu, \quad \frac{\partial \mathscr{O}_{ab}}{\partial \mathcal{R}^a_\mu} = (\boldsymbol{\mathcal{Q}}\transp \boldsymbol{\mathcal{Q}} \boldsymbol{\mathcal{R}}_b)_\mu,
\end{align}
while the $T$ derivatives act on $\boldsymbol{\mathcal{Q}}$ and require us to consider terms like
\begin{align}
    \sum_{\nu\tau \in \mathscr{E}} \mathcal{L}_\nu \frac{\partial \mathcal{Q}_{\nu\tau}}{\partial t_\mu} \mathcal{R}_\tau = \Tbraket{\mathcal{L}}{\tau_\mu \exp(T)}{\mathcal{R}},
\end{align}
where $\bra{\mathcal{L}}$ and $\ket{\mathcal{R}}$ denote general vectors.
The same terms arise in the case of SCCSD($\mathscr{K}$). Further details about the evaluation of these $T$ derivative terms are given in Supporting Information S5.

\section{Results and discussion}
The gradients and derivative coupling elements have been implemented in a development version of the $e^T$ program.\citep{eT2020} For comparisons of numerical and analytical gradients, see Supporting Information S6. Here, we will illustrate our implementation by locating minimum energy conical intersections (MECIs) for formaldimine and thymine and evaluating derivative coupling elements for lithium hydride. 

\subsection{Minimum energy conical intersections}
We apply the algorithm by Bearpark et al.\citep{BEARPARK1994269} to locate the MECIs. This algorithm constructs a gradient $\boldsymbol{G}$ that is zero when two conditions are fulfilled: (a) the energy difference between the two intersecting states vanishes, and (b) the energy gradient along the intersection seam is zero. In particular,
\begin{align}
    \label{eq:gradient-Bearpark}
    \boldsymbol{G} = \mathcal{P}_{gh^\perp} \nabla E_2 + 2 (E_2 - E_1) \frac{\boldsymbol{g}}{\vert\vert \boldsymbol{g} \vert\vert},
\end{align}
where
\begin{align}
    \boldsymbol{g} = \nabla (E_2 - E_1) = \boldsymbol{g}_2 - \boldsymbol{g}_1
\end{align}
and $\mathcal{P}_{gh^\perp}$ projects onto the complement of the $\boldsymbol{g}$-$\boldsymbol{h}$ plane, where
\begin{align}
    \boldsymbol{h} = (E_2 - E_1) \boldsymbol{d}_{12}.
\end{align}
Note that $\boldsymbol{g}$ and $\boldsymbol{h}$ define the directions along which the degeneracy is lifted.
In Eq.~\eqref{eq:gradient-Bearpark}, the second term in $\boldsymbol{G}$ vanishes when the energy difference between the states vanishes. The first term minimizes the energy of the upper surface along the seam by projecting out the components that lift the degeneracy (i.e., by projecting out the $\boldsymbol{g}$-$\boldsymbol{h}$ plane). The gradient $\boldsymbol{G}$ is used in combination with an existing Broyden-Fletcher-Goldfarb-Fanno (BFGS) algorithm.\citep{eT2020}

\subsubsection{Protonated formaldimine}
\begin{figure}[ht!]
    \centering
    \includegraphics[width=0.85\linewidth]{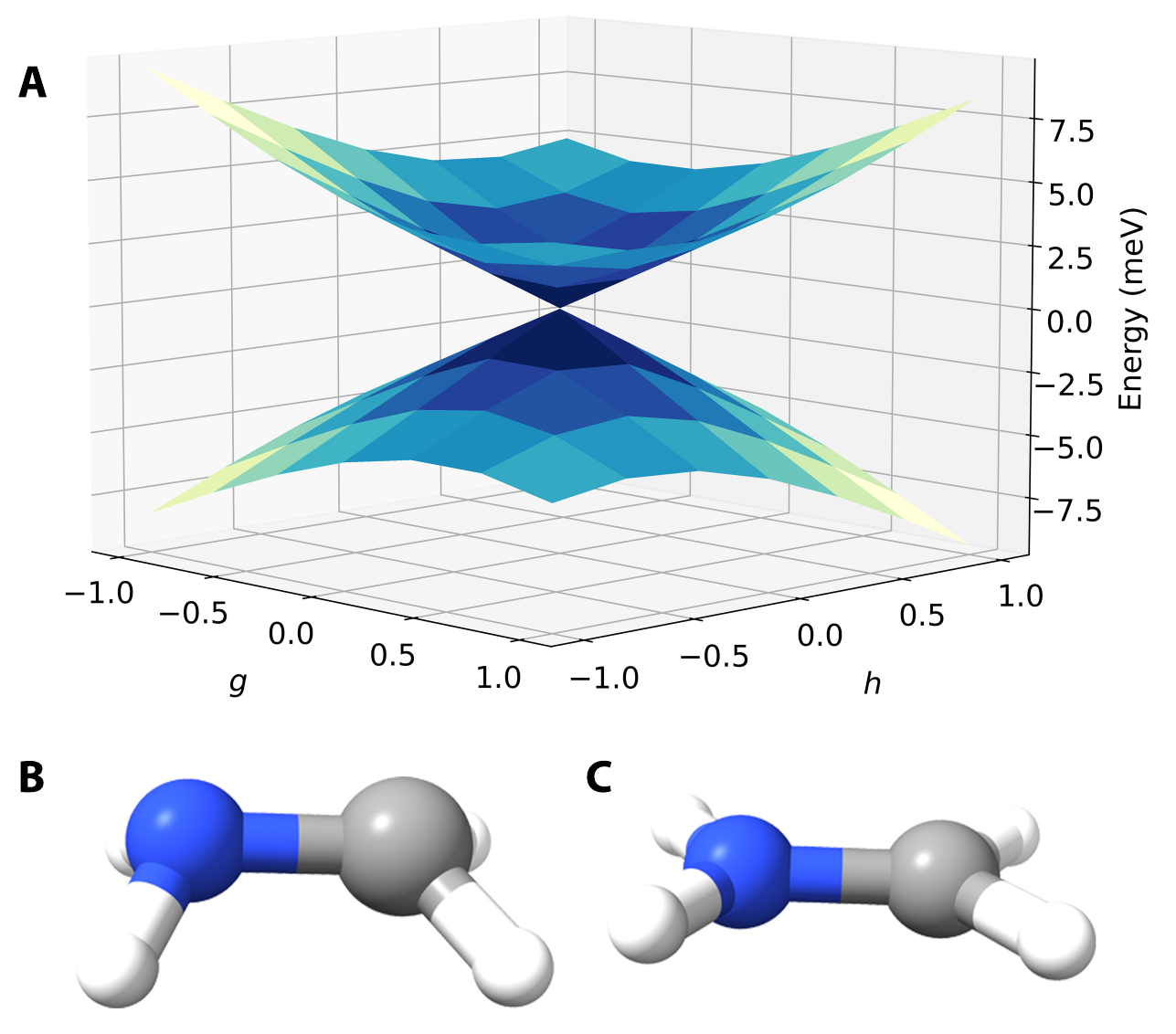}
    \caption{Formaldimine $S_1$/$S_2$ MECIs using SCCSD($\mathscr{E}$)/cc-pVTZ. \textbf{A}: Branching plane for distorted MECI. The $\boldsymbol{g}$ and $\boldsymbol{h}$ vectors are orthogonalized ($\boldsymbol{h}$ against $\boldsymbol{g}$) and given in arbitrary units; $\boldsymbol{h}$ has been rescaled to provide the same change in $S_1$/$S_2$ energy gap as displacements along $\boldsymbol{g}$. \textbf{B}, \textbf{C}: Distorted and planar MECI geometries, respectively.}
    \label{fig:formaldimine-meci}
\end{figure}

Table \ref{tab:MECI-formaldimine} and Figure \ref{fig:formaldimine-meci} show the optimized MECI geometries, along with a branching plane, for the first two excited states of protonated formaldimine ($S_1$ and $S_2$), the smallest model system for the chromophore in the light-sensitive protein rhodopsin.\citep{Auino2006}
We find two distinct MECIs, one that preserves the planar symmetry of the Franck-Condon geometry and one that breaks the planar symmetry (referred to as ``distorted''). The planar MECI was recently studied by Taylor et al.,\citep{taylor2023description} and here we compare our MECI geometries with theirs. Results for alternative projections are given in Supporting Information S7.

With SCCSD($\mathscr{E}$), we find that the $S_1$ and $S_2$ potential energy surfaces (see Figure \ref{fig:formaldimine-meci}A) have the correct double-cone topology in the vicinity of the $S_1$/$S_2$ intersection, as we have also observed previously in other systems.\citep{kjonstad2017resolving,kjonstad2019orbital,kjonstad2024thymine} Both the planar and distorted MECI geometries (see Figure \ref{fig:formaldimine-meci}B and C) are reached from the Franck-Condon geometry through an extension of the C-N bond. For the planar MECI, the bond extends from {$1.271$ Å} to {$1.426$ Å}. For this geometry, the CCSD and SCCSD methods are identical: the $S_1$ and $S_2$ states possess different symmetry, implying that the SCCSD correction is zero and the method reduces to CCSD. From the table, we note that the planar MECIs obtained with CCSD/SCCSD are in good agreement with the reference method in Ref.~\citenum{taylor2023description}, XMS-CASPT2. They also appear to be more accurate than the reported ADC(2) and TD-DFT geometries.

\begin{table}[ht!]
\centering
\begin{tabular}{lcccccc}
\hline
\hline
& CCSD & \multicolumn{2}{c}{SCCSD} & XMS-CASPT2 \cite{taylor2023description} & MP2/ADC(2) \cite{taylor2023description} & TD-DFT \cite{taylor2023description}\\
 & & ($\mathscr{E}$) & ($\mathscr{K}$) & \\
\hline
$S_0$ minimum & 1.271 & - & - & 1.281 & 1.275 & 1.274 \\
Planar MECI & 1.426& 1.426 &  1.426 & 1.420 & 1.389 & 1.541\\ 
Distorted MECI & 1.433  & 1.433  & 1.433 & - & -  & -\\ 	
\hline
\hline
\end{tabular}
\caption{C-N bond lengths (in $\si{\angstrom}$) in protonated formaldimine at the $S_0$ minimum and at $S_2$/$S_1$ minimum energy conical intersections. The TD-DFT calculations (from Ref.~\citenum{taylor2023description}) use the cc-pVDZ basis set. All other calculations are with the cc-pVTZ basis set.}
\label{tab:MECI-formaldimine}
\end{table}

If we allow the breaking of symmetry (by selecting an initial geometry that is non-planar), we find a distorted MECI with both CCSD and SCCSD that has a slightly longer C-N bond length of $1.433$ Å. In this case, the two states are also close to being of different symmetry since the distorted molecule has an approximate mirror plane. 

While these results show good agreement with literature,\citep{taylor2023description} we also find that the system illustrates some limitations of the method. In particular, because the orthogonality equation is non-linear in $\zeta$, it may have more than one solution. For formaldimine, considering a linear interpolation between the Franck-Condon geometry and the distorted MECI, we see that the solution that is well-behaved in the Franck-Condon region breaks down and is replaced by a different well-behaved solution at the MECI (see Supporting Information S8A).
This indicates that the SCCSD method must in some cases be considered a correction that should only be applied close to the intersection. This is not always the case, however, as the method has been successfully applied globally in dynamics simulations.\citep{kjonstad2024thymine} 
Careful considerations are needed when applying SCCSD locally in such simulations.

Dynamics simulations require smooth potential energy surfaces and derivative coupling elements, as this is necessary to avoid integration and energy conservation problems. Some care is therefore required when applying a trajectory-based dynamics method where SCCSD is switched on (instantly or, possibly, gradually) in the vicinity of conical intersections. In particular, an adaptive CCSD/SCCSD algorithm presupposes that the difference between CCSD and SCCSD is sufficiently small, and this is not guaranteed in general. However, it appears to be a reasonable assumption in many cases. In preliminary calculations on various systems (here and in other works\citep{kjonstad2017resolving,kjonstad2019orbital,kjonstad2024thymine}), we often find corrections on the order of $10^{-3}$ eV away from the defective intersections, with the onset of non-linearity in CCSD (coinciding with larger differences between CCSD and SCCSD) occurring only when the energy difference between the states is below $0.01$--$0.05$ eV. Moreover, in the context of dynamics simulations, corrections of this magnitude  resulted in practically identical time evolution with CCSD and SCCSD.\citep{kjonstad2024thymine} 
Nevertheless, a more detailed study is needed to reach general conclusions about such a CCSD/SCCSD algorithm.

\subsubsection{Thymine}
The thymine nucleobase efficiently relaxes back to the ground state after excitation by ultraviolet radiation, a property that has been linked to the resilience of genetic material to radiative damage.\cite{crespo2004} The first step in this non-radiative relaxation is believed to be an ultrafast (sub 100-fs) internal conversion from the bright $\pi\pi^\ast$ state ($S_2$) to a dark $n\pi^\ast$ state ($S_1$) through a conical intersection seam between these two states, as suggested by several theoretical and experimental studies.\citep{wolf2017probing, wolfguehr2019, mayer2024time, kjonstad2024thymine}

\begin{figure}[ht!]
    \centering
    \includegraphics[width=0.65\linewidth]{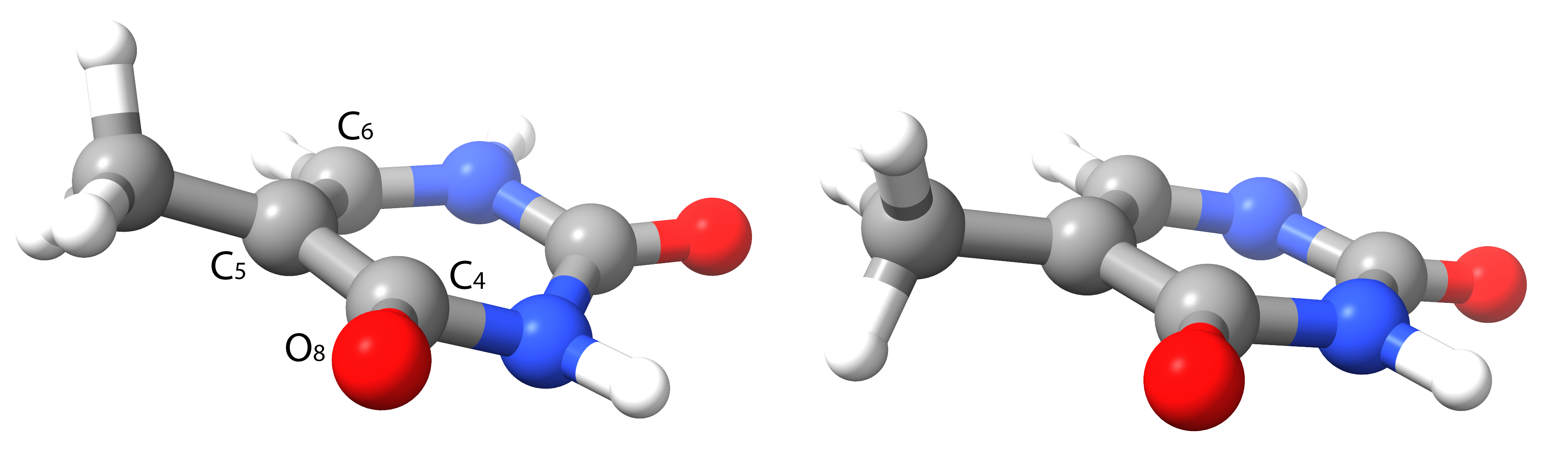}
    \caption{Thymine $S_1$/$S_2$ MECIs with SCCSD($\mathscr{E}$)/cc-pVDZ, distorted (left) and planar (right). Atomic labels for the C$_4$-O$_8$ and C$_5$-C$_6$ bonds are shown for the distorted MECI.}
    \label{fig:thymine-meci}
\end{figure}

We also find two MECIs for the nucleobase thymine, one that preserves the planar symmetry of the ground state geometry and one that is distorted and non-planar, see Table \ref{tab:thymine-meci} and Figure \ref{fig:thymine-meci}.  In the $\pi\pi^\ast$/$n\pi^\ast$ photorelaxation, two bond coordinates are believed to be particularly relevant:\citep{wolf2017probing} the C$_4$-O$_8$ and C$_5$-C$_6$ bonds (see Figure \ref{fig:thymine-meci}).
As the system moves away from the Franck-Condon region, it undergoes a long extension of the C$_5$-C$_6$ bond and a slight extension of the C$_4$-O$_8$ bond. The determined MECIs fit well with this picture. See Supporting Information S9 for MECIs obtained with other projections.

\begin{table}[ht!]
\centering
\begin{tabular}{lccccccccc}
\hline
\hline
& \hspace{1cm} $S_0$ minimum \hspace{1cm} & \multicolumn{3}{c}{Distorted MECI} &  & \multicolumn{3}{c}{Planar MECI} \\
& CCSD & CCSD & \multicolumn{2}{c}{SCCSD} & & CCSD & \multicolumn{2}{c}{SCCSD}\\
&  &  & ($\mathscr{E}$) & ($\mathscr{K}$) &  &  & ($\mathscr{E}$) & ($\mathscr{K}$)  \\
\hline
C$_4$-O$_8$ (Å) & 1.224 & 1.265*  & 1.265   & 1.265 & &  1.251  & 1.251   & 1.251 \\
C$_5$-C$_6$ (Å) & 1.357  & 1.446*  & 1.446   & 1.446 & & 1.465  & 1.465   & 1.465 \\
\hline
\hline
\end{tabular}
\caption{C$_4$-O$_8$ and C$_5$-C$_6$ bond lengths (in $\si{\angstrom}$) for thymine at the $S_2/S_1$ minimum energy conical intersections. The $S_0$ minimum was optimized using the aug-cc-pVDZ basis. All other calculations use the cc-pVDZ basis set. The asterisk indicates that the geometry was converged only to within $10^{-3}$ a.u.~in the gradient. Other geometries are converged to within $10^{-4}$ a.u.}
\label{tab:thymine-meci}
\end{table}

As for formaldimine, the planar MECIs in thymine are also identical with SCCSD and CCSD, and this is again because $S_1$ and $S_2$ possess different symmetries. In the non-planar case, we were not able to converge the CCSD MECI due to complex energies encountered during the optimization. Nevertheless, the partially converged MECI coincides with the SCCSD MECI, indicating that the CCSD and SCCSD methods are highly similar in this region of the intersection seam. 

\subsection{Lithium hydride}
As a final illustration, we present
derivative coupling elements for lithium hydride. This has been used as a test system of nonadiabatic couplings in CCSD\citep{Tajti2009,kjonstad2023coupling} and here we reconsider it for SCCSD. In previous works, the CCSD couplings have been shown to agree well with the full configuration interaction (FCI)
couplings, even in the dissociation limit. 

\begin{figure}[ht!]
    \centering
    \includegraphics[width=\linewidth]{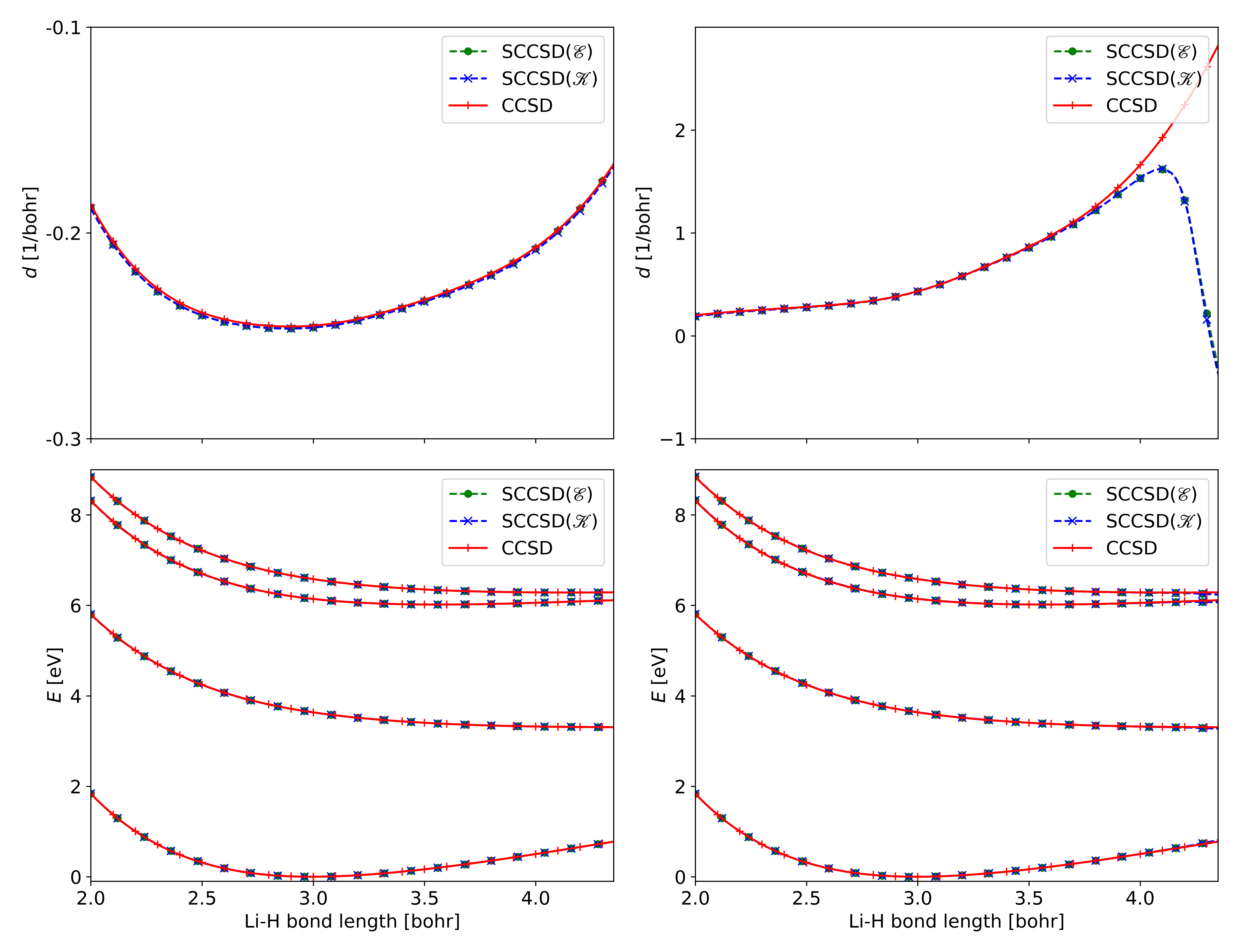}
    \caption{Coupling elements between the $2 \, ^{1}\Sigma^+$/$3 \, ^{1}\Sigma^+$ (top left) and $3 \, ^{1}\Sigma^+$/$4 \,^{1}\Sigma^+$ (top right) states. The magnitude of the coupling is calculated as $d = 2 (d_{\mathrm{Li}} - d_{\mathrm{H}})$. Energies of the considered states ($1 \, ^{1}\Sigma^+$, $2 \, ^{1}\Sigma^+$, $3 \, ^{1}\Sigma^+$, $4 \, ^{1}\Sigma^+$) are given in the bottom left and right panels. The couplings are the right coupling elements, that is, the nuclear derivative acts on the ket vectors (in this case, $3 \, ^{1}\Sigma^+$ and $4 \, ^{1}\Sigma^+$ for the left and right panels, respectively).}
    \label{fig:lih-1-4-4-5}
\end{figure}

In Figure \ref{fig:lih-1-4-4-5}, we compare couplings evaluated with standard and similarity constrained coupled cluster methods. Derivative couplings are shown for the $2 \, ^{1}\Sigma^+$/$3 \, ^{1}\Sigma^+$ and $3 \, ^{1}\Sigma^+$/$4 \,^{1}\Sigma^+$ states, where we consider bond distances near the equilibrium bond length. First, we find that the CCSD and SCCSD coupling elements are in close agreement close to the equilibrium (from $2.0$ to $4.0$ bohr), in most cases being so similar that the difference is invisible in the curves in the figure. However, the SCCSD coupling elements break down as the bond is extended beyond $4.0$ bohr for the $3 \, ^{1}\Sigma^+$/$4 \,^{1}\Sigma^+$ case (see Figure \ref{fig:lih-1-4-4-5}, right). We furthermore find that this breakdown is accompanied by a change in the well-behaved solution (see Supporting Information S8B). 

The breakdown in the dissociation limit is perhaps an unsurprising result, given that coupled cluster theory is often unreliable in this limit. Notably, however, the CCSD method correctly reproduces the nonadiabatic couplings both for the $2 \, ^{1}\Sigma^+$/$3 \, ^{1}\Sigma^+$ and the $3 \, ^{1}\Sigma^+$/$4 \,^{1}\Sigma^+$ 
states up to bond lengths as large as 8.0 bohr.\citep{Tajti2009,kjonstad2023coupling}  

\section{Conclusions}
In this paper, we have presented an implementation of analytical nuclear gradients and derivative coupling elements
for the similarity constrained coupled cluster singles and doubles method (SCCSD), building upon recent implementations for CCSD.\citep{schnackpetersen2022,kjonstad2023coupling} We have provided a few numerical examples, showing, for example, good agreement with literature values for a minimum energy conical intersection in protonated formaldimine, a simple model system for the chromophore in rhodopsin. However, we have also shown that the SCCSD method can in some cases have multiple solutions, suggesting that the method must in some cases (though not all, see Ref.~\citenum{kjonstad2024thymine}) 
be considered a local correction. 

Nevertheless, our implementation has made possible the first nonadiabatic dynamics simulations using both CCSD and SCCSD, as demonstrated in a separate nonadiabatic dynamics study on the ultrafast $\pi\pi^\ast$/$n\pi^\ast$ photorelaxation in thymine.\citep{kjonstad2024thymine} Given the accurate treatment of dynamical correlation in coupled cluster theory, we expect that its application in nonadiabatic dynamics simulations will provide valuable insights about the photochemistry of a variety of interesting systems.

\section*{Acknowledgements}
We thank Todd Martínez for discussions of the method and for hosting EFK as a visiting researcher in his research group, during which significant parts of the current work were developed and implemented. 
This work was further supported by the Norwegian Research Council through FRINATEK project 275506 and the European Research Council (ERC) under the European Union’s Horizon 2020 Research and Innovation Program (grant agreement No.~101020016).

\bibliography{paper}

\end{document}